\documentclass{ws-procs9x6}

\begin{document}
\title{The Discrete Nonlinear Schr\"odinger equation -- 20 Years on}
\author{J.~Chris Eilbeck}
\address{Department of Mathematics,\\
  Heriot-Watt University,\\
  Edinburgh EH14 4AS, UK\\
E-mail: J.C.Eilbeck@hw.ac.uk}
\author{Magnus Johansson}
\address{Dept of Physics and Measurement Techn.\\
  Link\"oping University,\\S-58183 Link\"oping, Sweden\\
E-mail: majoh@ifm.liu.se}

\maketitle 

\abstracts{We review work on the Discrete
  Nonlinear Schr\"odinger (DNLS) equation over the last two decades.}

\section{Introduction}
The Discrete Nonlinear Schr\"odinger (DNLS) equation describes a
particularly simple model for a lattice of coupled anharmonic
oscillators.  In one spatial dimension, the equation in its simplest
form is
\begin{equation}
 i{dA_j \over dt} +\gamma |A_j|^2A_j+ \varepsilon(A_{j+1} + A_{j-1}) = 0, 
\label{dnls}
\end{equation}
where $i=\sqrt{-1}$, the index $j$ ranges over the 1D lattice.  The
lattice may be infinite ($j=0, \pm1, \pm2, \dots$) or finite
($j=1,2,\dots,f$).  In the latter case one usually assumes periodic
boundary conditions, $A_{j+f}=A_j$.  The quantity $A_j=A_j(t)$ is the
complex mode amplitude of the oscillator at site $j$, and $\gamma$ is
a anharmonic parameter.

The connection with the continuous Nonlinear Schr\"odinger (NLS) equation
\begin{equation}
i A_t +\gamma |A|^2A +A_{xx}=0\label{nls}
\end{equation}
is more clear if we write (\ref{dnls}) in an alternative form
\begin{equation}
 i{dA_j \over dt} +\gamma |A_j|^2A_j+  \varepsilon(A_{j+1} -2A_j+
 A_{j-1}) = 0. 
\label{dnls1}
\end{equation}
The transformation $A_j\rightarrow A_j\exp(-2it\varepsilon)$ takes
(\ref{dnls1}) into (\ref{dnls}).  With $\varepsilon= 1/(\Delta x)^2$, 
(\ref{dnls1}) is seen as a standard finite difference 
approximation to (\ref{nls}).  

The DNLS equation (\ref{dnls}) is a special case of a more general
equation, the Discrete Self-Trapping (DST) equation\cite{els85} 
\begin{equation}
 i\frac{dA_j}{dt} + \gamma
|A_j|^2A_j+ \varepsilon\sum_{k}m_{jk}A_k =0.\label{dst}
\end{equation}
Here $M = [m_{ij}]$ is a $f \times f$ coupling matrix.  In physical
applications $M$ is real and symmetric, and clearly with $M$ a
suitably chosen constant tridiagonal matrix, we can regain (\ref{dnls})
or (\ref{dnls1}).  Physically this corresponds to the choice of
nearest neighbour couplings.  A more general choice for the elements
of $M$ introduce longer range couplings or different topologies to the
lattice.  The distinction between the DST and the DNLS equation is
somewhat blurred in modern usage.

One interesting limiting special case of the DST equation is the case
of the so-called {\em complete graph model}, when
$$
m_{ij} = 1-\delta_{ij},
$$
corresponding to a lattice with each point connected directly to every
other point on the lattice. 

Clearly one can scale $t$ and $\gamma$ in the DNLS model to fix
$\varepsilon=1$, and this is often done in the literature.  However
the more general formulation (\ref{dnls}) is useful when one wishes to
consider the case $\varepsilon\rightarrow0$, i.e.\ the limit of zero
coupling, known nowadays as the anti-integrable or anti-continuum 
limit\cite{ma94}. In this limit, the solution of (\ref{dnls}) is trivial:
\begin{equation}
A_j=\sqrt{\frac{\omega_j}{\gamma}} e^{i(\omega_j t + \alpha_j)},
\label{a-c}
\end{equation}
where the frequencies $\omega_j$ and phases $\alpha_j$ can be chosen 
arbitrarily and independently at each site.

It is worth pointing out to avoid confusion that there are other
possible discretizations of the NLS equation, one being the eponymous
Ablowitz-Ladik (AL) model\cite{al76a}
\begin{equation}
 i{dA_j \over dt} +\left(1+\frac12\gamma |A_j|^2\right)
(A_{j+1} + A_{j-1}) = 0.  \label{AL}
\end{equation}
Another is a model due to Izergin and Korepin\cite{ik81} which is 
rather lengthy to write down, see the book by Faddeev and
Takhtajan\cite{ft87} for details.  Both the AL model and the
Izergin--Korepin models have the advantage of being integrable
equations\cite{ac91}, but it can be argued that they are less
physically meaningful.  The DST equation has been shown to be
non-integrable when $f>2$ except for the rather unphysical case of a
special non-symmetric interaction matrix $M$.  It should also be
mentioned that an often studied model is the Salerno
equation\cite{sa92b}, which contains a parameter interpolating between
the AL model (\ref{AL}) with pure inter-site nonlinearity and the DNLS
equation (\ref{dnls}) with pure on-site nonlinearity. This allows
e.g.\ the use of soliton perturbation theory\cite{vg87,ckks93} to
elucidate the role of the on-site nonlinearity as a non-integrable
perturbation to the integrable AL equation.  The Salerno equation was
extensively analyzed by Hennig and co-workers\cite{dst201} and is also
described in the book by Scott\cite{sc99}.

Another possible source of confusion is that the acronym DNLS is
sometimes used for the Derivative Nonlinear Schr\"odinger
equation\cite{ac91}.

The DST equation (\ref{dst}) can be derived from the Hamiltonian:
\begin{equation}
 H = \sum^f_{j=1}\left[\omega_0^{(j)}  |A_j|^2-
{\gamma \over 2} |A_j|^4\right]
- \varepsilon \sum_{j \neq k}m_{jk} A_j^{*}A_k
\label{hamil}
\end{equation}
with canonical variables: $q_j \equiv A_j$ and $p_j \equiv {\rm
  i}A_j^{*}$.  Here $\omega_0^{(j)} \equiv \varepsilon m_{jj}$ are the
harmonic frequencies of each uncoupled oscillator ('on-site
energies'); with $m_{jk}=\delta_{j,k\pm1}$ the DNLS equations
(\ref{dnls}) and (\ref{dnls1}) are obtained for $\omega_0^{(j)}=0$ and
$\omega_0^{(j)}=2\varepsilon$, respectively.  There is a second
conserved quantity, the {\em number} (or {\em norm})
\begin{equation}
N = \sum^f_{j=1}|A_j|^2.
\label{norm}
\end{equation}
The integrability of the $f=2$ (dimer) case follows from these two
conserved quantities, and in this case all the time-dependent
solutions $A_j(t),j=1,2$ can be expressed in terms of elliptic
functions\cite{je82}.  One can always scale $A_j$ and $\gamma$ so that
$N=1$, or alternatively scale $A_j$ and $N$ so that $\gamma=1$.

The DNLS Hamiltonian is the starting point for a study of a quantum
version of the DNLS system, see the paper by Eilbeck in these
proceedings.  In particular the quantum analogue of a classical
discrete breather can be derived.
 
There are now almost 300 papers on the DNLS and DST equations, and in
this short survey we can only hope to cover a small amount of
available material, concentrating on our own interests.  Currently a
database of papers in this area is held at {\tt
  http://www.ma.hw.ac.uk/$\sim$chris/dst}. For complementary aspects
we recommend the review papers by Hennig and Tsironis\cite{dst201} (in
particular concerning map approaches with applications to stationary
wave transmission) and Kevrekidis et al.\cite{dstm256} (in particular
concerning different types of localized modes and their stability,
bifurcation and interaction properties), as well as the pedagogical
introduction in the textbook by Scott\cite{sc99} and the general
review of discrete breathers by Flach and Willis\cite{dst179}.

\section{Stationary Solutions}
\begin{figure}[th]
\centerline{\epsfxsize=3.5in\epsfbox{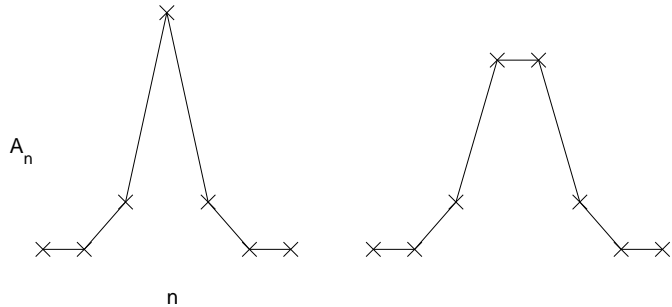}}   
\caption{Example: localized stationary solutions of the DNLS model
 \label{fig1}}
\end{figure}

Stationary solutions of the DNLS or DST equations are special
solutions of the form
\begin{equation}
A_j(t)=\phi_j \exp(i\omega t),
\label{stat}
\end{equation}
where the $\phi_j$ are independent of time.  Inserting this ansatz
into the equations give an algebraic set of equations for the
$\phi_j$.  For example, for DNLS (\ref{dnls}), we get
\begin{equation}
-\omega\phi_j +\gamma |\phi_j|^2\phi_j+ \varepsilon(\phi_{j+1} +
\phi_{j-1}) = 0.
\label{dnlsstat}
\end{equation}
It is this feature that makes the DNLS  a relatively simple model
to work with.  For small periodic lattices up to $f=4$ it is possible
to solve the resulting equations exactly and obtain all the families
of stationary solutions as a function of $\omega$ and $\gamma$ (for
fixed $N$), with a fascinating bifurcation structure\cite{els85}.
The complete graph model can also be solved exactly for any
$f$\cite{els85}.  For a large or infinite lattice the solutions must
be found by numerical methods such as shooting methods or spectral
methods.  These solutions can then be investigated as a function of
the parameters of the equation by numerical continuation methods (see
e.g.\ \cite{ei87} for a complete list of solutions for $f=6$).  If
$\gamma$ is sufficiently large, localized solutions are found which
decay exponentially for large $|j|$.  Two examples are
shown in Fig.\ref{fig1}.
Since these solutions have a periodic time behaviour $\phi_j
\exp(i\omega t)$, it seems appropriate to call them ``breather''
solutions.  Another motivation is that the DNLS equation can be
derived from the discrete Klein-Gordon equation, describing a lattice
of coupled anharmonic oscillators, via a multiscale expansion in the
limits of small-amplitude oscillations and weak inter-site
coupling\cite{kp92,ddp97,dstm260}.  The discrete breathers of this
lattice are then represented as stationary solutions to the DNLS
equation.  The reader should note that in the early days of DNLS
studies, when breathers in discrete systems were not so well
understood, these solutions were often called solitons.

The stability of such solutions in time can be investigated by looking
at general perturbations {\em in the rotating frame} of the
solutions\cite{ce85}
$$
\tilde A_j(t)=(\phi_j +\delta u_j(t))\exp(i\omega t)
$$
This reduces the linear stability problem to a study of a linear
eigenvalue problem.  It is perhaps to be expected that the stability
of a branch of stationary solutions can change at a bifurcation point.
What is surprising is that, since the eigenvalue problem is not
self-adjoint, solutions can also change stability at other points on
the branch.  Usually the single-site peaked ('site-centred') solution
shown at the l.h.s. of Fig.\ref{fig1} turns out to be stable, whereas
the two-site peaked ('bond-centred') solution shown on the r.h.s. is
not.

For the case of an infinite lattice, both solutions in Fig.\ref{fig1}
can be smoothly continued versus coupling $\varepsilon$ (or,
equivalently by rescaling, versus $\omega$) for all $\varepsilon$,
without encountering any bifurcations. For $\varepsilon \rightarrow
0$, the site-centred solution will be completely localized at the
central site with all other oscillator amplitudes zero, while the
bond-centred solution becomes completely localized on the two central
sites. For $\varepsilon\rightarrow\infty$ both solutions are smoothly
transformed into the same soliton solution of the continuous NLS
(\ref{nls}) (which explains why they are sometimes also termed
'discrete solitons'). As there are no bifurcations, the site-centred
solution is stable and the bond-centred unstable for all $\varepsilon$
in the infinite chain. Comparing the value of the Hamiltonian
('energy') of the two solutions for a fixed $N$, one finds that the
site-centred solution always has the lowest energy. This energy
difference has been proposed to act as a sort of Peierls-Nabarro potential
barrier\cite{dst89}. Another property of these two solutions in
infinite 1D lattices is that they exist for arbitrarily small $\gamma$
(or arbitrarily small $N$ for fixed $\gamma$)

{\bf Historical note}.  Although the stationary DNLS equation 
(\ref{dnlsstat}) was derived already by Holstein in 1959 in his 
pioneering work on polarons in molecular crystals\cite{ho59},  
the first systematic study of its single-peak breather solution 
as an exact solution to the fully discrete equations was performed
by Scott and MacNeil in 1983\cite{sm83},
following Scott's interest in Davydov solitons on proteins.  They
investigated the family of single-peak stationary solutions using
shooting methods running on a Hewlett-Packard programmable calculator.
Further interest in Davydov solitons on protein molecules led to a
study of a related molecule called {\em acetanilide}.  A model of the
crystalline state of this molecule was set up which was essentially
four {\em coupled} DNLS systems.  Techniques to map out the families
of stationary solutions on this system were developed, including
path-following from the anharmonic limit\cite{els84}.  It was then
realised that the {\em single} DNLS system was of independent
interest, which led to the work described in\cite{els85}. 

Later and independently, Aubry, MacKay and co-workers developed a much
more general approach along these lines to the breather problem in
arbitrary systems of coupled oscillators\cite{ma94,dst157}.  In this
context, much new attention was directed to the DNLS model and its
stationary solutions.  In addition, two more large bursts of interest
into studies of the DNLS equation have appeared recently, following
the experimental progress in the fields of nonlinear optical waveguide
arrays\cite{esmba98} and Bose-Einstein condensates trapped in periodic
potentials arising from optical standing waves\cite{ak98}. These
applications will be discussed briefly below. Since the DNLS equation
is of general applicability and appears in completely different
physical fields, new researchers drawn to its study have not always
been aware of earlier results. Thus many of its properties have
been independently rediscovered and appeared several times in the
literature in different contexts during the last two decades.
 
\section{Disorder}
One natural generalization to the DNLS equation (\ref{dnls}) or
(\ref{dnls1}) is to consider non-constant coupling parameters
$\varepsilon_{jk}$, equivalent to nontrivial distributions of the
elements $m_{jk}$ of the matrix $M$ in the DST equation (\ref{dst}).
One may also consider site dependent $\gamma_j$ as well.  An early
application, in the large $f$ case, was to model the dynamics of the
energy distribution of modes on a globular protein.
Feddersen\cite{fe91a} considered interactions among CO stretch
oscillations in adenylate kinase, which comprises 194 amino acids ($f
= 194$).  Since the structure of this enzyme has been determined by
x-ray analysis, the $f(f-1)/2 = 18721$ off-diagonal elements of the
dispersion matrix $M$ were calculated from Maxwell's equations. Also
diagonal elements were selected from a random distribution, and the
degree of localization of a particular stationary solution of the form
(\ref{stat}) with real $\phi$ was defined by evaluating the quotient $
\sum \phi_j^4 / \sum \phi_j^2$.

This numerical study revealed two features. Firstly, at experimentally
reasonable levels of nonlinearity ($\gamma$), stable localized
solutions were observed near some but not all of the lattice sites.
Secondly, this anharmonic localization was observed to be distinctly
different from ``Anderson localization'', a property of randomly
interacting linear systems \cite{an78}. Thus none of the stationary
states that were observed to be highly localized at large $\gamma$
remained so as $\gamma$ was made small. Also, none of the states that
were localized at $\gamma = 0$ (i.e., Anderson localized) remained so
as $\gamma$ was increased to a physically reasonable level.

The transition between Anderson localized modes and breather states
has more recently been extensively analysed in a series of papers by
Kopidakis and Aubry for general coupled oscillator
chains\cite{ka99,ka00,ka00prl} (see also Archilla et al.\cite{amm99}
for a slightly different model), and has to its larger parts been
understood.  The generic scenario, valid also for the DNLS model, is
consistent with Feddersen's observations but too complicated to
describe in detail here. Briefly, there are two kinds of localized
breather solutions in a disordered nonlinear lattice: `extraband
discrete breathers' (EDBs) with frequencies outside the spectrum of
linear Anderson modes, and `intraband discrete breathers' (IDBs) with
frequencies inside the linear spectrum. EDBs cannot be smoothly
continued versus frequency into IDBs but are lost in cascades of
bifurcations. IDBs on the other hand can be continued outside the
linear spectrum, but not into EDBs but only into a certain type of
spatially extended multi-site breathers. The IDBs can only exist as
localized solutions inside the linear spectrum provided their
frequencies do not resonate with linear Anderson modes.  However, for
an infinite system the linear spectrum becomes dense so that the
allowed frequencies for localized IDBs must constitute a (fat, i.e. of
non-zero measure) Cantor set! In fact, for the DNLS case the latter
result had been rigorously obtained already in 1988 by Albanese and
Fr{\"o}hlich\cite{af88}; see also\cite{fsw86,af88II} for other early
mathematical results on the DNLS model with disorder.

It is interesting to remark that the general scenario with two types
of discrete breathers, EDBs and IDBs, where the latter exist as
localized single-peaked solutions only in-between resonances with
linear modes, is not peculiar for random systems, but observed also in
other situations when the linear spectrum is discrete with localized
eigenmodes. A very recently studied example\cite{ppl02} is the DNLS
model (\ref{dnls}) with an added linearly varying on-site potential
$\omega_0^{(j)} = \alpha j$. In this case the linear spectrum
constitutes a so-called Wannier-Stark ladder (WSL) of equally spaced
eigenfrequencies, with eigenstates localized around each lattice site,
giving rise to Bloch oscillations (recently experimentally observed in
waveguide arrays\cite{mpaes99,pdebl99}).  Then, resonances were shown
to result in 'hybrid discrete solitons', interpreted as bound states
of single-peaked IDBs and satellite tails corresponding to nonlinearly
modified Wannier-Stark states localized some distance away from the
main peak. Due to the finite (constant) distance between the linear
eigenfrequencies in the WSL, IDBs remain single-peaked in frequency
intervals of finite length, in contrast to the IDBs in random systems.

The fact that nonlinearity modifies localized linear modes into
extended nonlinear solutions should be of some physical importance,
since it provides a mechanism for transport in random systems. In
fact, this aspect was considered also by Shepelyansky in
1993\cite{sh93}, who used the well-known Chirikov criterion of
overlapping resonances to argue that, above some critical nonlinearity
strength $\gamma_c$, the number of excited linear modes (and thus the
spatial width) for a typical initially localized excitation in the
DNLS model would spread sub-diffusively as $(\Delta n)^2 \sim t^{2/5}$
(for linear random systems, $(\Delta n)^2$ remains finite under very
general conditions).

We also mention that the case with disorder residing purely in the
nonlinearity strengths $\gamma_j$ was studied by Molina and
Tsironis\cite{dst102}. In this case only partial localization of an
initially single-site localized excitation could be found for large
nonlinearities (dynamical self-trapping, corresponding to asymptotic
approach to an exact discrete breather), while some portion was found
to always escape ballistically (i.e. spreading as $(\Delta n)^2 \sim
t^2$) leading to absence of complete localization. The scenario with
partial self-trapping above a critical nonlinearity strength combined
with asymptotic spreading through small-amplitude waves appears very
generally for single-site initial conditions in the DNLS model, with
or without disorder\cite{dst77,dst84,jhr95}.

\section{Mobile breathers}
Since the DNLS equation is an approximation to the equations
describing Davydov solitons, which are thought from numerical studies
to be mobile, it is natural to ask whether the sort of breathers shown
in Fig.\ref{fig1} can move if sufficiently perturbed.  The first
attempt to model this was made in a relatively obscure conference
proceedings\cite{ei86} in 1986.  The key to getting mobility is to
realise that a shape like Fig.\ref{fig1} will move if the figure
represents $|A_j|^2$ but the phase is no longer constant and rotates
through $2\pi$ as we traverse the breather.  The same paper reported a
very preliminary study of the interaction of the moving breather with
an ``impurity'', or more precisely a long-range interaction due to the
curved nature of the chain.  There is now a growing literature on
trapping of mobile breathers due to curved chains and long
range-coupling (c.f.\cite{fe91a,dst238}) and on trapping due to local
impurities in the lattice (c.f.\cite{dst132}).  Regrettably, due to
space considerations, we have omitted any further discussion of this
interesting area.

Feddersen\cite{fe91,defw93} used spectral and path-following methods to
make a more detailed numerical study of travelling breathers in the
DNLS system using the ansatz
\begin{equation}
A_j(t)=u(z) \exp\{i(\omega t-\kappa j)\}, \quad z=j-ct.
\label{travel}
\end{equation}
Note that $c \ne \omega/\kappa$, i.e.\ the solution is regarded as a
solitary pulse modulated by a carrier wave moving at a different
velocity.  His studies show a solution with this form to a high
degree of numerical accuracy for a range of parameter values.  However
this numerical evidence cannot be regarded as a rigorous proof for the
{\em existence} of moving breathers in the DNLS system, and this is
still an outstanding question. 

Much recent attention has been drawn to mobile breathers in general
oscillator chains (see several other contributions to these
proceedings), and many of these results can be transferred also to
DNLS chains.  Here, we wish to just mention particularly some results
of Flach et al.\cite{fzk99} (see also \cite{fk99}) who used an inverse
approach, choosing particular given profiles of travelling waves and
finding equations of motion having these as exact solutions.
Generalizing the DNLS equation (\ref{dnls}) by replacing $\gamma
|A_j|^2$ with $G(|A_j|^2)$ and $\varepsilon$ with $\varepsilon +
F(|A_j|^2)$, where $F$ and $G$ are functions to be determined, and
choosing $A_j$ of the form (\ref{travel}) with {\em real} $u$, they
could determine explicit expressions for the functions $F$ and $G$ for
which the particular solution $A_j$ exists as an exact travelling
wave.  In this way, they could e.g. reproduce the AL-model (\ref{AL})
for $G\equiv 0$ by choosing $A_j$ to be its well-known soliton
solution. Moreover, they could prove that no such travelling solution
with pulse shaped $u$ could exist for a pure on-site nonlinearity
($F\equiv 0$), and thus not for the DNLS equation (\ref{dnls}).
However, this does not prove the absence of exact moving localized
DNLS breathers, for (at least) two reasons.  (i) The envelope $u(z)$
could contain a non-trivial space-dependent complex phase not
absorbable into the $\exp\{-i\kappa j\}$ factor. In fact, the
solutions numerically found by Feddersen contained such a phase. (ii)
Moving breathers could have a time-varying (e.g.\ periodic) shape
function $u(z,t)$. This could be possible since the stationary DNLS
breathers, in the regime where mobility is numerically observed,
exhibit internal shape modes ('breathing modes') which can be found as
localized time-periodic solutions to the linearized
equations\cite{dstm192,kpcp98,dst233,dstm256}.

\section{Chaotic Solutions}
Since the DNLS equation for $f>2$ is not integrable, it might be
expected that it has solutions exhibiting Hamiltonian chaos, and in
fact the first study of the DST equation showed chaotic-looking
trajectories in the $f=3$ case\cite{els85}.  A more thorough analysis
of this case was carried out by Cruzeiro-Hansson et
al.\cite{cffcss90}, who estimated the region of both classical and
quantum phase space occupied by chaotic states.  A number of other
studies have been carried out since then. For example, Hennig and
coworkers\cite{dstm122} considered a DST trimer (\ref{dst}) with
$m_{11}=m_{22}=m_{33}=0$ and $m_{13}=m_{23} \ll m_{12} =m_{21}$, i.e.\ 
an (integrable) dimer interacting weakly with the third oscillator.
Then, a Melnikov approach could be used to show the existence of
homoclinic chaos. A similar approach for the case $f=4$ with a dimer
interacting weakly with the two other sites also demonstrated the
presence of Arnold diffusion\cite{dst126}. In the opposite limit of
large $f$, homoclinic chaos has also been demonstrated and analyzed
through a Melnikov analysis of a perturbed continuous NLS
equation\cite{cems96}.

As another example of chaotic behaviour, Eilbeck\cite{ei87,es89}
showed that on a $f=6$ periodic lattice modelling benzene, a mobile
breather could propagate which hopped around the lattice in a random
way, even reversing its direction of motion at unpredictable
intervals.

\section{2-dimensional DNLS lattices}
As follows from the general theory of MacKay and Aubry\cite{ma94},
breathers exist also in higher dimensions. While we are aware of very
few explicit results for the DNLS model in three
dimensions\cite{fkm97}, the two-dimensional case has been rather
thoroughly studied. Some recent results are described
in\cite{dstm256}.  Instead of attempting to give a complete survey
here, we will concentrate on discussing the main differences to the
one-dimensional case.

In the 2D case and for a square lattice, the DNLS equation
(\ref{dnls1}) with $\gamma=\varepsilon=1$ is readily generalized to
\begin{equation}
 i{dA_{m,n} \over dt} + |A_{m,n}|^2A_{m,n}+ 
A_{m+1,n} + A_{m-1,n} + A_{m,n+1} + A_{m,n-1}-4A_{m,n}  = 0,
\label{dnls2d}
\end{equation}
and stationary solutions of the form (\ref{stat}) with frequency
$\omega$ can be found analogously to the 1D case. The single-site
peaked discrete soliton (breather) was first thoroughly studied
in\cite{dst120,dst133}. The following characteristics should be
mentioned: (i) The solution can be smoothly continued from a
single-site solution at the anti-continuum (large-amplitude) limit
$\omega\rightarrow \infty$ to the so called ground state solution of
the continuous 2D NLS equation\cite{rr86} in the small-amplitude limit
$\omega\rightarrow 0$.  (ii) There is an instability-threshold at
$\omega \sim 1$, so that the solution is stable for larger $\omega$
('discrete branch') and unstable for smaller $\omega$ ('continuum-like
branch'). (iii) The stability change is characterized by a change of
slope in the dependence $N(\omega)$, so that ${dN \over d\omega} > 0$
($<0$) on the stable (unstable) branch. (A similar criterion exists
also for single-site peaked solutions to the 1D DNLS equation with
on-site nonlinearities of arbitrary power\cite{dst105}.)  (iv) The
value of the excitation number $N$ at the minimum is {\em nonzero},
and thus there is an {\em excitation threshold}\cite{dst210} for its
creation, in contrast to the 1D case (\ref{dnls1}) where $N\rightarrow
0$ as $\omega\rightarrow 0$ for fixed $\gamma=\epsilon=1$. A similar
scenario occurs also in 3D\cite{fkm97}. The effect of this excitation
threshold in 2D was recently proposed to be experimentally observable
in terms of a delocalizing transition of Bose-Einstein condensates in
optical lattices\cite{dstm268}. (v) The dynamics resulting from the
instability on the unstable branch is, in the initial stage, similar
to the collapse of the unstable ground state solution of the
continuous 2D NLS equation\cite{rr86}, with increased localization and
blow-up of the central peak. In contrast to the continuum case,
however, this process must be interrupted since the peak amplitude
must remain finite, and the result is a highly localized 'pulson'
state where the peak intensity $|A_{m,n}|^2$ oscillates between the
central site and its four nearest neighbours\cite{dstm144}. This
process has been termed 'quasicollapse'.\cite{dst105,dst120} It is not
known whether these pulson states represent true quasiperiodic
solutions to the DNLS equation (see below).

As was shown by MacKay and Aubry\cite{ma94,dst157} under very general
conditions, two-dimensional lattices allow for a new type of localized
solutions, 'vortex-breathers', with no counterpart in 1D. They can be
constructed as multi-site breathers by continuation from an
anti-continuum limit of a cluster of single-site breathers with
identical frequencies but with uniformly spatially varying phases
constituting a closed loop, such that the total phase variation around
the loop ('topological charge') is a multiple of $2\pi$. The simplest
examples are three breathers in a triangle phase shifted by $2\pi/3$, or
four breathers in a square phase shifted by $\pi/2$. The general
existence and stability theorems\cite{ma94,dst157} (valid also for the
DNLS equation) guarantee that such solutions exist as localized
solutions for weak enough coupling, and that certain configurations
are linearly stable.  As a consequence of the phase torsion, such
solutions will carry a localized circulating current when the coupling
is nonzero.  For the DNLS model, vortex-breathers for a square 2D
lattice were first obtained in \cite{dstm192}.  Typically they become
unstable as the coupling is increased; the mechanisms of these
instabilities were described in some detail in\cite{mk01,dstm256}.

Let us finally also mention a recent study\cite{dstm270} exploring
numerically different types of breathers (including vortex-breathers)
and their stability in triangular and hexagonal DNLS-lattices.

\section{Quasiperiodic Breathers}
A particular feature of the DNLS equation, distinguishing it from
generic anharmonic lattice models, is the existence of continuous
families of exact, spatially localized solutions of the form
(\ref{stat}) but where the amplitudes $\phi_j$ in the rotating frame
are not time-independent but time-periodic (with nonharmonic
time-dependence). Such solutions are obtained by adding a term ${i
  \dot{\phi}_j}$ to the left-hand side of Eq.(\ref{dnlsstat}).  Thus,
these solutions are in general {\em quasiperiodic} with two
incommensurate frequencies in the original amplitudes $A_j$ (although
they may also be periodic if the frequency relation is rational). At
first, one may not be surprised by the existence of quasiperiodic
solutions, since at least for finite-size lattices they should appear
as KAM tori. However, generically (i) one would not expect them to
appear in continuous families since they should be destroyed for
rational frequency relations; and (ii) one would not expect them to
survive as localized solutions in infinite lattices since the presence
of two incommensurate frequencies in a generic anharmonic system would
generate all possible linear combinations of the frequencies, i.e. a
dense spectrum implying that resonance with the continuous spectrum
should be unavoidable, and the breather should radiate and decay.

The key point to realize why, in spite of this, quasiperiodic
breathers with two incommensurate frequencies do exist in the DNLS
lattice is to note that the first frequency $\omega$ in (\ref{stat})
always yields harmonic oscillations, and thus no multiples of this
frequency are generated.  The origin to this is the {\em phase
  invariance} of the DNLS equation, i.e.\ invariance under
transformations $A_j \rightarrow A_j e^{i\alpha}$, related to the norm
conservation law (\ref{norm}) by Noether's theorem.  A recent
result\cite{bv02} proves that very generally, each conservation law in
addition to the Hamiltonian yields possibility for existence of
quasiperiodic breathers with one additional frequency.

The existence of quasiperiodic DNLS-breathers in infinite lattices was
first proposed by MacKay and Aubry\cite{ma94}, and later explicit
proofs of existence and stability as well as numerical demonstrations
for some particular examples were given\cite{dst167,dstm192} (earlier
findings of quasiperiodic solutions in DNLS-related models had
concerned mainly small systems\cite{je82,dst76} or the integrable AL
model\cite{cbg95}). As some renewed interest has appeared on this
topic\cite{dstm256}, it is useful to comment on the differences
between these two approaches. The solutions in\cite{dst167,dstm192}
(see also \cite{bv02}) were constructed as multi-site breathers by
continuation from the anti-continuum limit $\varepsilon=0$ of
solutions with two (or more) sites oscillating with non-zero amplitude
according to (\ref{a-c}) with two different (generally incommensurate)
frequencies $\omega_1$ and $\omega_2$. Except for some particular
relations between the frequencies where resonances appear, such
solutions can always be continued to some non-zero $\varepsilon$. On
the other hand, the solutions discussed in\cite{dstm256} originated in
internal-mode excitations (i.e.\ time-periodic localized solutions to
the linearized equations) of a particular stationary solution, the
so-called 'twisted localized mode'\cite{dst189} (TLM). As for the
bond-centred breather in Fig.\ref{fig1}, the anti-continuum version of
this solution has two neighbouring sites oscillating with equal
$|A_j|^2$; however for the TLM these sites are oscillating in
anti-phase so that the solution is spatially antisymmetric. This
solution exists and is linearly stable for small
$\varepsilon$.\cite{dst167} Now, the occurrence of linear
internal-mode oscillations is a very common phenomenon\cite{kpcp98}.
However, in most cases such oscillations do not yield true
quasiperiodic solutions of the fully nonlinear equations since
typically some harmonic will resonate with the linear continuous
phonon spectrum, implying that these oscillations decay in time. This
scenario appears e.g.\ for the single-site peaked
DNLS-breather\cite{dst233}. The interesting discovery by Kevrekidis
and co-workers was, that for the particular case of the TLM, there are
certain intervals in $\varepsilon$ where {\em all }higher harmonics of
the internal mode frequency are outside the continuous spectrum, and
thus in these intervals the oscillating solutions of the linearized
equations could be continued into truly quasiperiodic localized
solutions of the nonlinear equation. As the allowed intervals are away
from $\varepsilon=0$, it is clear that this approach yields solutions
which could not be obtained by direct continuation from the
anti-continuum limit.

\section{Wave Instabilities}
Another important class of solutions in anharmonic lattices are space-time 
periodic {\em travelling waves}. For the DNLS model such solutions are 
very simple, since they are just rotating-wave solutions of the type 
(\ref{travel}) with constant $u=|A|$. Direct substitution into the DNLS 
equation (using the form (\ref{dnls1})) yields the nonlinear dispersion 
relation 
\begin{equation}
\omega= - 4\varepsilon \sin^2{\frac{\kappa}{2}} + \gamma |A|^2. 
\label{disp}
\end{equation}
Linear stability analysis shows\cite{ce85,kp92} that the travelling waves are 
stable if and only if $\frac{\gamma}{\epsilon} \cos \kappa < 0$. Thus, for 
$\frac{\gamma}{\epsilon}>0$ only waves with $\pi/2<|\kappa|\leq\pi$ are 
stable, while waves with small wave vectors $0\leq|\kappa|\leq\pi/2$ are 
unstable through a {\em modulational instability} analogously to the 
continuous NLS equation. This instability destroys the homogeneous amplitude 
distribution of the wave, and typically\cite{ddp97,rabt00,rckg00} 
results in the creation of a number of small-amplitude 
mobile localized excitations ('breathers'), which through interaction 
processes (see below) may coalesce into a small number of standing 
large-amplitude breathers. Thus, the plane-wave modulational instability 
was proposed\cite{ddp97} generally to constitute the first step towards 
energy localization in nonlinear lattices (including DNLS).

Now, in a linear system one may always take linear combinations of
counter-propagating waves $e^{\pm i \kappa j}$ to obtain {\em standing
  waves} (SWs) of the form $\cos(\kappa j + \beta)$. The same is of
course not true in a nonlinear system due to lack of superposition
principle; still however there generally exist nonlinear continuations
of the linear standing waves, although they cannot be written as
superpositions of counter-propagating travelling waves. Such nonlinear
standing waves were investigated in detail for general coupled
oscillator chains in \cite{dstm260} (the results for the DNLS chain
were more concisely summarized in \cite{jmak02}). Without going into
too much detail, let us state some main conclusions, referring to the
DNLS form (\ref{dnls1}) with $\gamma=1$. (i) SWs with given wave
vector $\kappa$ exist as stationary solutions of the form (\ref{stat})
for {\em all }values of $\frac{\omega}{\varepsilon} > - 4
\sin^2{\frac{\kappa}{2}}$.  In the lower limit (corresponding to the
dispersion relation (\ref{disp}) for a linear wave), the wave is a
linear standing wave. (ii) In the anti-continuum limit
$\frac{\omega}{\epsilon}\rightarrow\infty$ a SW with wave vector
$\kappa$ is described by a particular spatially periodic (or
quasiperiodic if $\kappa$ is irrational) repetition of local on-site
solutions of the form (\ref{a-c}) of oscillating and zero-amplitude
solutions. The oscillating sites have the same frequency $\omega$ but
generally alternating phases $\alpha=0,\pi$.  (iii) For each wave
vector $\kappa$ there are only two different distinct (modulo lattice
translations) SW families corresponding to different spatial phases
$\beta$ of the linear SW $\cos(\kappa j + \beta)$. They appear as
hyperbolic respectively elliptic periodic points in the map defined by
the stationary DNLS equation (\ref{dnlsstat}). (iv) One of the SW
families is stable in a regime of large $\frac{\omega}{\epsilon}$,
while close to the linear limit {\em all nonlinear SWs with
  $\kappa\neq\pi$ are unstable} for infinite systems! The instability
for the 'most stable' waves is of oscillatory type (i.e. corresponding
to complex eigenvalues of the linear stability eigenvalue problem).

Investigating the long-time dynamics resulting from the SW
instabilities, completely different scenarios were found\cite{jmak02}
for $|\kappa|<\pi/2$ and $|\kappa|>\pi/2$, respectively. For the first
case one finds after long times persisting large-amplitude standing
breathers, while for the second case a 'normal' thermalized state is
obtained. In fact, this division of the available phase space into two
isolated regimes of qualitatively different asymptotic dynamics was
first found by Rasmussen et al.\cite{rckg00}, and shown to correspond
to a phase transition through a discontinuity in the partition
function in the Gibbsian formalism. In terms of the Hamiltonian and
norm densities for a chain of $f$ sites, 
the phase transition line was obtained as $\frac{H}{f}
= -\gamma (\frac{N}{f})^2$, which can be seen to correspond exactly to
a SW with wave vector $|\kappa| = \pi/2$. Note that the existence of
the second conserved quantity $N$, which is peculiar for DNLS-type
models, is crucial in this context.

Another interesting observation is that taking the limit
$\kappa\rightarrow\pi$ for one of the nonlinear SWs generated from the
anti-continuum limit as above, one obtains a solution consisting of a
stable background wave with $\kappa=\pi$ having a single defect site
of zero-amplitude oscillation inserted into it. This solution can be
smoothly continued to the continuum limit, where it is seen to
correspond to the dark-soliton solution of the {\em defocusing} NLS
equation (note that the transformation $A_j=(-1)^j A_j$ in (\ref{dnls}) is
equivalent to reversing the sign of $\frac{\gamma}{\varepsilon}$).
Also the discrete dark soliton ('dark breather') has been shown to be
stable close to the anti-continuum limit, but unstable through an
oscillatory instability close to the continuum limit for arbitrarily
weak discreteness\cite{jk99}.  The typical outcome of this instability
is a spontaneous motion\cite{dst116}. As for the case of ordinary
moving breathers, it is still an open question whether moving dark
breathers exist as exact solutions, and current research is devoted to
this issue. However, numerical evidence that they can exist at least
to a very high numerical accuracy was given in the work of
Feddersen\cite{fe91,fe93}.

Let us also mention that {\em asymmetric} discrete dark solitons, with 
different left and right background amplitudes, can exist as quasiperiodic 
solutions of the type described in the previous section. Such solutions were 
analyzed by Darmanyan et al\cite{dkl01} and are subject to similar 
instabilities.

\section{Breather Interactions}
In general, one cannot conclude from a linear stability analysis that a 
solution is fully stable, but only that small perturbations at least 
cannot grow exponentially in time. However, for the single-peaked 
DNLS-breather, a stronger result is obtained\cite{dst210}: 
such solutions are orbitally Lyapunov stable for {\em norm-conserving} 
perturbations. This basically means that small breather perturbations 
will remain small (modulo a possible phase drift) for {\em all} times. 
This result is a consequence of the single-site breather being a ground 
state solution, in the sense that among all possible solutions at a 
given norm, it has the smallest value of the Hamiltonian. Thus, once again 
we find a property where norm conservation is crucial, and thus one should 
not expect that Lyapunov stability is a generic property of breathers 
in Hamiltonian lattices. 

Still there are important issues to address concerning the fate of perturbed 
breathers, which cannot be predicted from stability theorems. One issue 
is breather-breather interactions, which correspond to large 
perturbations. Some preliminary work was done by Feddersen\cite{fe91},
who showed that the collison of two breathers of equal amplitude
travelling in opposite directions was close to elastic.  In the more
general case  the situation is more complicated.

Accumulated knowledge from several numerical experiments on 
general breather-carrying systems, in particular by Peyrard and coworkers 
(e.g.\cite{pey98}) has lead to the conjecture, that in collisions between 
standing large-amplitude breathers and moving small-amplitude breathers, 
{\em big breathers systematically grow at expense of the small ones}. For 
the DNLS model, such a scenario was described in \cite{rabt00}.

Another interesting issue is breather interactions with
small-amplitude phonons, where also the long-time dynamics cannot be
predicted from stability theorems since extended phonons in infinite
lattices have infinite norm. A first approach\cite{lk00,kk00} is to
consider this as a linear scattering problem, with incoming, outgoing
and reflected linear phonons scattered by the breather. Then, within
the linear framework, one finds the scattering on a single-peaked
DNLS-breather to be always elastic. In certain cases, even perfect
transmission or perfect reflection of phonons appear\cite{lk00,kk00}.

However, going beyond linear theory the scattering process is
generally inelastic, and the breather may absorb or emit energy to the
surrounding phonons. These processes were investigated in
\cite{dst233,jo01} using a multiscale perturbational approach. It was
found\cite{dst233}, that under certain conditions a breather can pump
energy from a single phonon and continuously {\em grow} with a linear
growth rate.  This process is always associated with generation of
second-harmonic outgoing phonon radiation. On the other hand, it was
also found\cite{jo01} that breather {\em decay} could only happen if
two or more different phonons were initially simultaneously present.
An additional interesting observation\cite{jo01} was that beyond a
certain breather amplitude ($|A_0|^2\gtrsim 5.65$ corresponding to
$\omega>4\epsilon$ for the DNLS of form (\ref{dnls1}) with
$\gamma=1$), all lowest-order growth and decay processes disappear.
Thus, this explains why, once created, breathers with large amplitude
are extremely stable also for non-norm-conserving perturbations.

Let us finally mention also some results obtained\cite{dstm156} for an
extended DNLS model, which has very recently received renewed
attention in the description of ultrafast catalytic electron
transfer\cite{ak02}. To model the interaction of an electron, or
exciton, with a classical phonon system treated as a thermal bath, the
DNLS equation is appended with the terms $\left[-\eta{d \over dt}
  (|\psi_j|^2)+h_j(t)\right]\psi_j$, where the first term is a
nonlinear damping term providing dissipation, and the second term is a
fluctuation term which as a crudest approximation is taken as a
Gaussian white noise. This extended DNLS equation conserves excitation
number but not the Hamiltonian. Then, it was shown\cite{dstm156} that
breathers are always ultimately destroyed, but that strongly localized
breathers may be very long-lived for weak noise. The decay was shown
to be linear in time, with decay rate proportional to
$D\left({\varepsilon/\gamma}\right)^2$, where $D$ is the noise
variance (here $N=1$ is assumed). It would be highly interesting to
know whether similar behaviour could appear also in more realistic
models with coloured noise, since the white noise can be considered to
be somewhat unphysical having infinite frequency content.

\section{Applications}
We have already mentioned the Holstein polaron model as (to our
knowledge) the first\cite{ho59} suggested application of a DNLS
equation.  Likewise, we mentioned Davydov
solitons\cite{dav73,sc99,dst59}.  Another early motivation for the
study of the DNLS/DST equation was within the theory of Local Modes of
small molecules\cite{se86b}. The two latter topics are well
described in the textbook by Scott\cite{sc99}.
Here we  just briefly discuss the two applications which have
attracted the most attention during the last five years, namely
coupled optical wave guides and Bose-Einstein condensates (BEC).

The modelling of two coupled optical waveguides, interacting through 
a nonlinear material, by a DNLS dimer equation was suggested already in 1982 
by Jensen\cite{je82}. Later work\cite{cj88} extended these ideas and 
proposed the DNLS equation to describe discrete self-focusing in arrays 
of coupled waveguides. Many works followed proposing the applicability 
of different properties of the DNLS equation for nonlinear optical 
purposes; here we just mention the investigation of packing, steering and 
collision properties of self-localized beams\cite{dst134}, and the use 
of discreteness effects to obtain a controlled switching between different 
guides in the array\cite{bm96}. The success of the DNLS equation in 
describing discrete spatial solitons in waveguide arrays was first 
experimentally confirmed in 1998.\cite{esmba98} Later experimental work 
showed the existence of propagating discrete solitons and confirmed the 
DNLS predictions of a Peierls-Nabarro barrier\cite{mpaes99a} as well as 
that of nonlinear Bloch oscillations\cite{mpaes99}. More recently, also 
dark discrete solitons were observed\cite{messa01}. 

In the context of Bose-Einstein physics, the use of the dimer DNLS
equation was (to our knowledge) first suggested by Smerzi et
al.\cite{sfgs97} to model two weakly coupled BEC in a double-well
trap. Later\cite{ts01}, the full DNLS equation was proposed to model
the earlier quoted experiment\cite{ak98} with a BEC trapped in a
periodic potential, and the existence of discrete solitons/breathers
for such experiments was predicted. A large amount of theoretical
predictions, based on DNLS dynamics, for different phenomena to occur
in BEC arrays has appeared in the last year, of which we here, quite
randomly, just quote \cite{abdks01,dstm268}. So far, most of the
predictions are awaiting experimental verification. Some experimental
confirmation that, at least to some extent, BEC in periodic potentials
can be treated with DNLS models, under the condition that the
inter-well potential is much larger than the chemical potential, has
appeared very recently\cite{cat01,cat02}. In these experiments the BEC
was trapped in an optical lattice superimposed on a harmonic magnetic
potential, and modelled by a DNLS equation with an additional
quadratic on-site term $\Omega j^2 A_j$. The observed frequency of the
Josephson-like coherent oscillations of the BEC centre-of-mass in the
magnetic trap was shown to agree with DNLS predictions\cite{cat01}.
Moreover, changing the centre of the magnetic potential led to a
transition from the (superfluid) regime of coherent oscillations into
an insulator regime with the condensate pinned around the potential
centre\cite{cat02}. The onset of the transition was interpreted as the
result of a discrete modulational instability, and could be estimated
from the DNLS model. Many new experiments in this exciting field are
awaited in the near future!

\section{Conclusions}
We hope the reader has enjoyed this brief introduction to this
fascinating topic.  We are conscious of the many details, figures
and areas that we have left out, either because of space restrictions
or because the topics are covered in depth elsewhere.  To do the
subject full justice would require a whole volume.

\section*{Acknowledgements}
We thank all our colleagues, too many to mention explicitly, which in one 
way or another have contributed to this field. 
Special thanks are due to Sergej Flach and Thomas Pertsch for their 
helpful assistance, 
and to Rolf Riklund and Michael {\"O}ster for reading the manuscript.
JCE would like to thank the EU for the financial support of the LOCNET
programme.  MJ would like to thank the Swedish Research Council for support.

\bibliography{dst.bib,dst1.bib,dstm.bib}
\bibliographystyle{plain}
\end{document}